\documentclass[twocolumn,superscriptaddress,showpacs,amsmath,amssymb,prl,longbibliography,floatfix]{revtex4-1}

\usepackage[pdftex]{graphicx} 
\usepackage{dcolumn}  
\usepackage{bm}       
\usepackage[usenames,dvipsnames]{color}
\definecolor{URLCOL}{rgb}{0,0.52,0.83} 
\definecolor{LINKCOL}{rgb}{0.05,0.5,0} 
\definecolor{orange}{rgb}{0.6,0.3,0} 
\definecolor{CITECOL}{rgb}{0.25,0,0.48} 
\usepackage{epstopdf}

\usepackage[pdftex,bookmarks,breaklinks,bookmarksopen,bookmarksnumbered,colorlinks,linkcolor=LINKCOL,linktocpage,citecolor=CITECOL,urlcolor=URLCOL,pdfpagemode=UseOutline,pdftex]{hyperref}

\usepackage{soul}
\usepackage{booktabs}
\usepackage{natbib}

\definecolor{TITLECOL}{rgb}{0.1,0.2,0.7} 
\definecolor{SECOL}{rgb}{0.1,0.2,0.7} 
\definecolor{CONTENTSCOL}{rgb}{0.1,0.2,0.7} 
\definecolor{SSECOL}{rgb}{0.25,0,0.48} 
\definecolor{SSSECOL}{rgb}{0.2,0.08,0.53} 
\definecolor{FINCOL}{rgb}{0.01,0.3,0.07} 

\def\coloredtitle#1{\title{\textcolor{TITLECOL}{#1}}} 
\def\coloredauthor#1{\author{\textcolor{CITECOL}{#1}}} 

\definecolor{URLCOL}{rgb}{0,0.17,0.43} 
\definecolor{LINKCOL}{rgb}{0.05,0.4,0} 
\definecolor{CITECOL}{rgb}{0.35,0,0.48} 

\def\sss{\scriptscriptstyle\rm}

\def\bea{\begin{eqnarray}}
\def\eea{\end{eqnarray}}
\def\ben{\begin{equation}}
\def\een{\end{equation}}
\def\benu{\begin{enumerate}}
\def\enu{\end{enumerate}}

\def\bei{\begin{itemize}}
\def\eei{\end{itemize}}
\def\beit{\begin{itemize}}
\def\eit{\end{itemize}}
\def\benu{\begin{enumerate}}
\def\enu{\end{enumerate}}

\def\br{{\bf r}}

\def\half{\frac{1}{2}}

\def\s{_{\sss S}}

\def\szero{_{{\sss S},0}}

\def\xc{_{\sss XC}}

\def\Hxc{_{\sss HXC}}
\def\Hxczero{_{{\sss HXC},0}}
\def\xczero{_{{\sss XC},0}}
\def\H{_{\sss H}}

\def\intr{\int d^3r\,}

\def\n{n}

\def\dv{\Delta v}

\def\sec#1{\section{\textcolor{SECOL}{#1}}}
\def\ssec#1{\subsection{\textcolor{SSECOL}{#1}}}

\def\taur{{\tau_R}}

\def\Hxctau{_{{\sss Hxc},\tau}}
\def\n{n}
\def\tauapp{{\tau,\rm{app}}}

\def\t{^\tau}

\begin{document}
\sf

\coloredtitle{
Thermal stitching:  Extending the reach of quantum fermion solvers
}
\coloredauthor{Justin C. Smith}
\affiliation{Department of Mathematics, University of California, Los Angeles, CA 90095}
\affiliation{Department of Physics and Astronomy, University of California, Irvine, CA 92697}
\coloredauthor{Kieron Burke}
\affiliation{Department of Physics and Astronomy, University of California, Irvine, CA 92697}
\affiliation{Department of Chemistry, University of California, Irvine, CA 92697}

\date{\today}

\begin{abstract}
For quantum fermion problems, many accurate solvers are limited by the
temperature regime in which they can be usefully applied.
The Mermin theorem implies the uniqueness of an effective potential from 
which both the exact density and free energy at a
target temperature can
be found, via a calculation at a different, reference temperature.
We derive exact expressions for both the potential and the free energy in such
a calculation, and introduce three
controllable approximations that reduce the cost of such calculations.
We illustrate the effective potential and its free energy, and test
the approximations, on the asymmetric two-site Hubbard model at finite temperature.
\end{abstract}

\maketitle
\def\sec#1{}
\def\ssec#1{}

\sec{Introduction}

The fermionic quantum problem occurs in many areas of physics and
is notoriously difficult to solve\cite{D29}.  
It is at the heart of all electronic structure problems,
and so solution methods have enormous impact in condensed matter physics,
quantum chemistry, materials science, and beyond\cite{PGB14}.  Over decades, many
diverse approaches have been developed and refined\cite{MRC16}.  In almost all
cases, there are trade-offs between accuracy, computational cost, and
domain of applicability.  Some techniques are almost solely designed
to work on finite systems at zero temperature (e.g., many ab initio
quantum chemical approaches), while others are extremely general but
costs become prohibitive as the temperature lowers (e.g., path integral
Monte Carlo, PIMC\cite{Cc95,DM12}).  A collection of high-accuracy methods has recently been
benchmarked on small strongly-correlated lattice models\cite{MCCG17}.  On the other hand
density-functional methods are relatively inexpensive, but 
require an uncontrolled approximation to the exchange-correlation (XC)
energy.  Recently, density functional theory (DFT) methods have enjoyed considerable success in being
applied at temperatures relevant to warm dense matter (WDM), a phase of matter
with properties between those of solids and plasmas\cite{GDRT14}, such
as occurs in
fusion experiments and planetary cores\cite{RAYS01,HRRR04,GLNL07,SGGV08,KRDM08,LHR09,DOE09,EHHS09,MBRK09,BBFH10,RMCH10,SEJD14,FLDG15}.

The central question addressed in this work is:  Is there some way
that a quantum fermion solver could be run at one
temperature (the reference temperature) to yield results at some other
temperature (the target temperature)?  Such a scheme could be applied
to many diverse combinations of calculations.  In the examples above,
it could be used to bootstrap PIMC calculations to lower temperatures,
quantum chemical calculations to finite temperatures, or to combine
DFT methods with more accurate solvers for WDM\cite{M09b,DM15,SGVB15}.

We show that the answer is in principle yes,
at least for extracting the free energy and density at the target temperature.
Inspired by Ref. \cite{DP00}, we use the Mermin theorem\cite{M65} to define a unique effective
one-body potential from which, with an accurate quantum solver, we can extract the target quantities.  
We derive the relevant formulas for a finite-temperature Kohn-Sham
treatment.  We identify three useful, controllable approximations that 
make extraction of the target free energy easier in practice.  Finally,
we illustrate the relevant exact quantities and test the approximations 
on the finite-temperature asymmetric Hubbard dimer.

Although our formulas might be applied to any quantum fermion problem, we will 
discuss WDM simulations as a concrete example.
Mermin generalized the Hohenberg-Kohn theorem\cite{HK64}
to systems in thermal equilibrium at non-zero temperatures\cite{M65}. Thermal DFT (thDFT) became
a popular tool of plasma physics in subsequent decades\cite{DT81,DP82,PD84}.
The advent of accurate ground-state approximations and robust materials codes led to
many recent successes
of thDFT\cite{MD06,LHR09,KDBL15,KD09,KDP15,HRD08,KRDM08,RMCH10,SEJD14,WHMG17}.
For greater reliability and in principle higher accuracy, but at much higher computational
cost, 
path integral Monte Carlo (PIMC) simulations are 
used\cite{Cc95,MC00,FBEF01,SBFH11,DM12,DGSM16,DGMS17}.

As in the ground state, thDFT is made computationally tractable by the use of a 
non-interacting potential $v\s\t(\br)$
that yields the interacting density, $n\t(\br)$, at temperature
$\tau$. This Mermin-Kohn-Sham 
(MKS) system is exact in principle but in practice
requires approximations to the 
exchange-correlation (XC) free energy, $A\xc^\tau[\n]$, as a functional of the
density\cite{KS65}.
Most WDM simulations use
the zero-temperature approximation (ZTA)\cite{SPB16}, in which
$A\xc\t[\n]$ is replaced by $E\xc[\n]$, an approximation to the ground state
XC energy\cite{MOB12}, but used in the thermal MKS equations.
An alternative is to use the thermal local density approximation,
where a parametrization of the XC free energy
of the homogeneous electron gas is used to approximate
$A\xc^\tau[n]$\cite{KSDT14,DGSM16,GDB17}. 
Thermal generalized gradient approximations\cite{SD14,KDT17} have also been suggested.

\sec{Theory}

\ssec{Notation}

\newcommand{\denpot}[1]{\mathcal{I}\left[ #1 \right]}
\newcommand{\calI}[1]{\mathcal{I}\left[ #1 \right]}

Begin with the Mermin-Kohn-Sham scheme.  The equations are 
identical to those of the ground state,
\ben
\left\{-\half \nabla^2 + v\s\t(\br) \right\}\phi_i\t(\br) = \epsilon_i\t \, \phi\t_i(\br),\\
\label{MKSeqn}
\een
with the exception that the density is found by thermally occupying the MKS orbitals:
\ben
\n\t(\br) = \sum_i f_i\t |\phi\t_i(\br)|^2,
\label{MKSdens}
\een
where the occupations are Fermi factors at temperature $\tau$.  
$v\s(\br)$ is defined by Eqs. (\ref{MKSeqn}) and (\ref{MKSdens}).
Write the free energy in terms of the MKS components:
\ben
A[v]=\min_{\n} \left( T\s\t[\n] - \tau S\s\t[\n]+ U[\n]+ A\xc\t[\n] + \calI{\n v} \right),
\label{Av}
\een
where $T\s\t$ is the MKS non-interacting kinetic energy at temperature $\tau$,
$S\s\t$ is the corresponding entropy, $U$ is the Hartree energy, and we have
introduced
\ben
\calI{f} = \intr f(\br)
\een
to represent the external potential energy.
Writing 
\ben
v\s\t(\br)=v(\br)+v\H[\n](\br)+v\xc\t[\n](\br),
\label{MKSpot}
\een
and identifying $v\H(\br)$ as the Hartree potential and $v\xc\t(\br)$ as the
functional derivative of $A\xc\t$, the
self-consistent solution of the MKS equations finds the minimum
density in Eq. (\ref{Av}). 

\begin{figure}[htb]
\includegraphics[width=\columnwidth]{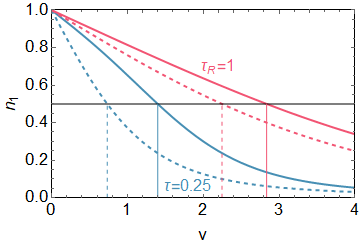}
\caption{
$n_1$ vs. $x$ at $\tau=0.25$ (blue) and $\tau=1$ (red). Solid lines are $U=1$ and dashed are non-interacting, $U=0$. 
The intersections with the horizontal line at $n_1=0.5$ give the $v$ values
that yield $n_1=0.5$ for the given temperature and interaction.
}
\label{stitchcon}
\end{figure}
We make this more concrete with a simple model.
The asymmetric Hubbard dimer has seen increasing use as an exact model to test and understand
many flavors of DFT including ground state\cite{CFSB15,CM16}, 
time-dependent\cite{FFTA13,FM14,FMb14,TR14}, ensemble\cite{DMF17}, 
thermal\cite{SPB16,BSGP16}, and DFT-like methods\cite{KSPB16}.  Its Hamiltonian is
\ben
\hat{H}=-t\sum_\sigma(\hat{c}^\dagger_{1\sigma}\hat{c}_{2\sigma}+\textrm{H.c.})
+\sum_i(U\hat{n}_{i\uparrow}\hat{n}_{i\downarrow}+v_i\hat{n}_i)
\een
where $\hat{c}^\dagger_{i\sigma}(\hat{c}_{i\sigma})$ is the electron creation (annihilation) operator
and $\hat{n}_{i\sigma}=\hat{c}^\dagger_{i\sigma}\hat{c}_{i\sigma}$ is the number operator,
$t$ is the strength of electron hopping, $U$ is the Coulomb repulsion, and $v_i$ is the onsite
potential. We choose $v_1 + v_2 =0$, define $v = v_2 - v_1$, and $2\,t=1$.
In lattice DFT the site-occupations\cite{SGN95}, $n_1$ and $n_2$, are the analogs of the density. 
We work at half-filling by setting $\langle N \rangle =2$ which restricts $\mu=U/2$ to
maintain particle-hole symmetry.
Fig. \ref{stitchcon} shows exact thermal calculations.  The solid red line is the
density on site 1 as a function of the onsite potential $v$, for a relatively hot
temperature ($\tau=1$).   The Mermin theorem guarantees its monotonicity.  
The dashed red line is the same map but for tight-binding, i.e., $U=0$.  Thus,
for a system with $v=2.834$ (marked by solid red vertical line),
$n=0.5$ at $\tau=1$.  The MKS potential is $v\s\t=2.246$ (vertical dashed red line),
and the difference is the HXC contribution.   The 
blue lines denote the same things at a lower temperature, $\tau=1/4$.

The logical basis of the MKS scheme is the Mermin theorem.
Mermin proved that in the grand canonical ensemble for fixed temperature
and chemical potential there exists a one-to-one correspondence between the external
potential and electronic density for given particle statistics, interaction, and
temperature, $\tau$\cite{M65}.  Assuming $v$-representability,
the map $\bar\n\t[v](\br)$ is invertible 
and the map
$v\t[\n](\br)$ exists.  Note that the former is a potential functional (denoted
by a bar),
while the latter is a density functional.
Assuming non-interacting representability,
we may then write
\ben
\bar v\s\t[v](\br)=(\n\s\t)^{-1}[\bar \n\t[v]](\br).
\label{KSinversion}
\een
This compact expression is the map between the one-body potential
of the interacting problem and its KS alter ego, i.e., this is
the KS potential as a functional of the one-body potential of the
interacting problem, which is different from
its density dependence as expressed in Eq. (\ref{MKSpot}).

\def\vtil{\bar{v}_\tau^\taur}
\def\Delv{\Delta {v}_\tau^\taur}
For any system, we can define an effective thermal potential,
$\vtil(\br)$, as the one-body potential that yields the exact density
at $\tau$ by performing a calculation
at $\taur$.  This is unique by Mermin's theorem and can be written
\ben
\vtil[v](\br) = \left(\n^\taur\right)^{-1}[\n^\tau[v]](\br).
\een
A non-interacting map is defined in the same way.
Figure \ref{stitchcon} also illustrates the effective thermal potential logic.
The horizontal line is $n_1=0.5$ and everywhere that it intersects a curve corresponds to the
potential that yields $n_1=0.5$ for the given temperature and interaction strength. 
Thus $\vtil[v]$ is given by the dependence of the blue vertical line on the red one,
with an analogous non-interacting version with dashed vertical lines.
This effective potential has some specific symmetry properties, namely
\ben
{\bar{v}_\taur^\tau}[\vtil[v]](\br)= {\bar{v}_\tau^\tau}[v](\br)=v(\br).
\label{inversion}
\een

\def\nt{\n\t\{\tilde v\xc\}[v]}
We now wish to derive the effective thermal potential for an MKS calculation, using some
$\tilde v\xc\t[n](\br)$, where this XC potential could be either approximate or exact.
To do this, we must use the concept of a ffunctional\cite{CGB13}.  A functional is a function of
a function, whereas a ffunctional is a functional of a functional.  Identify
$\n\t\{\tilde v\xc\}[v](\br)$ as the density at temperature $\tau$ found by solving the
MKS equations with $\tilde v\xc\t[n](\br)$. Then
\ben
\vtil[v](\br)= v\s^\taur[\nt](\br)-v\Hxc^\taur[\nt](\br).
\label{vtil}
\een
This result shows
how to construct
an approximate effective thermal potential from a MKS calculation
at two different temperatures with a given XC free energy potential.
Simply calculate the density from MKS at the desired temperature, find
what non-interacting potential yields that density at the reference temperature,
and subtract off the approximate HXC potential evaluated at the reference temperature.
In Fig. \ref{stitchcon}, the first term is the MKS contribution (vertical
dashed blue line), while the second is the HXC correction (difference
between solid and dashed vertical blue lines).
Thus a DFT approximation might be used to generate PIMC-quality densities 
at $\tau$
by performing
PIMC calculations only at $\taur$.
Our result satisfies several important conditions:
(i) if the exact XC functional is used, the exact $\n\t(\br)$ is found;
(ii) if an approximate XC is used, and the resulting effective thermal 
potential is used in a MKS calculation with the same XC approximation, the
corresponding self-consistent approximate density is found; (iii) if the
temperatures are set equal, the exact result is recovered.  But the
symmetry of Eq. (\ref{inversion}) is lost with an approximate XC.

\begin{figure}[htb]
\includegraphics[width=\columnwidth]{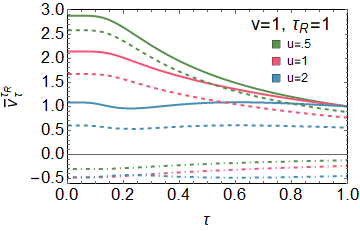}
\caption{
Effective thermal potential for $v=1$ and $\taur=1$ for various correlation strengths.
Solid curves are interacting,
dashed are non-interacting, and dot-dashed is Hxc. 
All calculations yield $\bar{v}\Hxctau^\taur \leq 0$.
}
\label{barstitchpot}
\end{figure}
In Fig. \ref{barstitchpot}, we plot the exact effective thermal
potential, for a system with $v=1$.  Green denotes weak correlation
($u=0.5$).  (In the ground state for $v=0$, the radius of convergence of
the small-$u$ expansion is 2).  The solid line is the interacting curve, which
varies strongly with temperature (but approaches $v$ as $\tau\to\taur$, as
required).  The dashed line is the MKS effective thermal potential, which mimics the interacting curve closely, and 
approaches the MKS potential at $\taur$.  The dot-dashed line is
the HXC contribution, which is relatively small and much smoother, suggesting
it might be amenable to approximation.

We also show what happens as we increase the correlation to $u=1$ (red) and
$u=2$ (blue).  For moderate correlation, the effects are similar, but larger.
But for strong correlation temperature dependence is mitigated, the HXC contribution is comparable to the MKS
piece, and small errors in its approximation are less likely to be forgiven.

\begin{figure}[htb]
\includegraphics[width=\columnwidth]{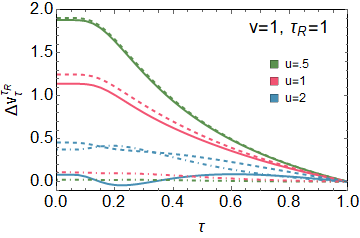}
\caption{
Same as previous figure, but now for the difference between effective thermal potential
and its reference.}
\label{stitchpot}
\end{figure}
In Fig. \ref{stitchpot},  we plot 
the difference between the effective thermal potential and 
its reference value ($v$ and $v\s\t$ for interacting and non-interacting,
respectively), showing that the HXC contributions are now even smaller.  They remain
monotonic when correlation is weak or moderate, and vanish rapidly as $\tau\to\taur$. 
This strongly suggests that approximating the XC contribution to the thermal
correction potential with a local or semilocal density functional approximation
should introduce relatively little error in the density for weakly correlated
systems.
For strong correlation, the HXC contribution is of the same
order as the MKS potential, develops nonmonotonic behavior, 
and vanishes much more slowly with temperature.  A semilocal density approximation
might introduce much larger errors in this case.

Although the density is important,
greatest interest is often in the free energy and related properties.
Thus we need to generate accurate free energies from our formulas.
We begin with a recently proven formula\cite{CP15} in potential functional
theory (PFT)\cite{YAW04,CLEB11,CGB13} to calculate the free
energy. In Ref. \cite{CP15} PFT is generalized to the grand canonical
ensemble.   Assume the energy components are known exactly for some given reference
potential, $v_0$, and write
$v^\lambda(\br) = v_0(\br) + \lambda \dv(\br)$, where $\dv(\br)=v(\br)-v_0(\br)$.
The free energy of the system is then:
\ben
A\t[v] = A\t_0 + \calI{\bar{n}\t[v,\dv] \dv},\label{CPeq}
\een
where
$\bar{n}^\tau[v,\dv] = \int_0^1 d\lambda\, n^\tau[v^\lambda](\br)$.
Here 0 subscripts denote quantities for the reference potential.
We find, exactly, for the deviation from the reference $A\Hxc\t$:
\begin{equation}
\begin{split}
\Delta A\Hxc\t[\dv] &= \calI{\bar{n}\t[v,\dv]\dv}
-\calI{\bar{n}\s\t[v\s\t,\dv\s\t]\dv\s\t} \\
&+  \calI{n\t[v]v\Hxc\t-n\t_0[v_0] v\Hxczero\t}.
\label{CPeqXC}
\end{split}
\end{equation}
The derivation of Eq. (\ref{CPeqXC}) is given in the
Supplemental Material\footnote{See Supplemental Material at [URL will be
inserted by publisher] for a thorough derivation
of the XC free energy equations and numerical demonstrations of the
various approximations.}.

\begin{figure}[htb]
\includegraphics[width=\columnwidth]{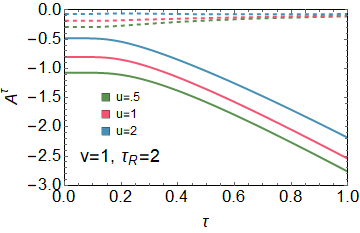}
\caption{
Temperature dependence of the free energy and its deviation from reference for the same systems
as in the previous figures. }
\label{freeenergy}
\end{figure}
To illustrate the value of a well-chosen reference, 
in Fig. \ref{freeenergy}, we plot the free energy versus temperature 
using Eq. (\ref{CPeq}) with the reference potential set to $0$, i.e., the perfectly
symmetric case.  We see that the deviation from the reference is an order
of magnitude smaller than the reference value, making it easier to approximate.
Note that here our reference temperature is twice as high as before, but even
at half its value, the deviation in the free energy from the reference is difficult
to detect.

In principle, Eq. (\ref{CPeqXC}) is sufficient to extract the free energy from 
a thermal-stitching calculation.  Although the input densities are required
at the target temperature $\tau$, these can all be found from calculations at
the reference temperature.  
The last term in Eq. (\ref{CPeqXC}) is straightforward,
but the first involves averages over $\lambda$ that are cumbersome since
the effective thermal potential must be evaluated for every $\lambda$.  
The last step in our work is to derive a controlled approximation that yields
an accurate expression using only quantities evaluated at $\taur$.

We make three distinct approximations.
In the first,
we note that the exact formula requires finding 
$\n\s\t[v\szero+\lambda\dv\s](\br)$ which, in general, 
is not equal to $\n\t[v_0+\lambda\dv](\br)$.
However, they match at $\lambda=0$ and $\lambda=1$, and nearly
agree everywhere for weak interaction,
so we expect 
\ben
\n\s\t[v\szero+\lambda\dv\s](\br)\approx\n\t[v_0+\lambda\dv](\br)
\een
to produce very little error.   A second approximation is to approximate each
coupling-constant integral by a two-point formula:
\ben
\bar{n}^\tau[v,\dv](\br) \approx \half (n^\tau[v_0](\br) + n^\tau[v](\br)). 
\label{approx2point}
\een
With these Eq. (\ref{CPeqXC}) greatly simplifies to
\begin{equation}
\begin{split}
\Delta A\xc^\tauapp[v]
&=
\calI{\left(n^\tau[v] - \bar{n}^\tau[v]\right) v^\tau\xc]} \\
&-\calI{\left(n^\tau[v_0] - \bar{n}^\tau[v]\right)  v\xczero^\tau },
\label{XCapprox}
\end{split}
\end{equation}
with the Hartree contributions explicitly canceling on both sides (See 
Supplemental Material for derivation\cite{Note1}.).
Inserting the effective thermal potential is now simple:
\begin{equation}
\begin{split}
\!\Delta A\xc^\tauapp[v] &=
\calI{(n^\taur[\tilde{v}_\tau^\taur[v]] - \bar{n}^\taur[\tilde{v}^\taur_\tau[v]]) v^\tau\xc} \\
&-\calI{(n^\taur[\tilde{v}_\tau^\taur[v_0]] - \bar{n}^\taur[\tilde{v}_\tau^\taur[v]])  v\xczero^\tau}.  
\label{XCcorrections} 
\end{split}
\end{equation}
This formula yields (approximately) the XC free energy at $\tau$ using only densities
from $\taur$, effective thermal potentials, and the XC potential at $\tau$,
which 
can be extracted via a MKS inversion from the accurate
density at $\tau$, i.e. Eq. (\ref{KSinversion}),
and subtraction of the external and Hartree potentials.

Although Eq. (\ref{XCcorrections}) contains only quantities evaluated at the
reference temperature, as required, they are awkward because the reference
potentials and densities must be found for many values of $\lambda$, and then
averaged over the coupling constant.  This process can be simplified by
a linear approximation for the thermal effective potential:
\bea
\tilde{v}^\taur_\tau[v^\lambda](\br) &=&
\tilde{v}^\taur_\tau[v_0 + \lambda (v-v_0) ](\br) \nonumber \\
&\approx& \tilde{v}^\taur_\tau[v_0](\br) +
\lambda (\tilde{v}^\taur_\tau[v](\br) - \tilde{v}^\taur_\tau[v_0](\br)),
\label{lambdaapprox}
\eea
which should be an excellent approximation for weak correlation.

\begin{figure}[htb]
\includegraphics[width=\columnwidth]{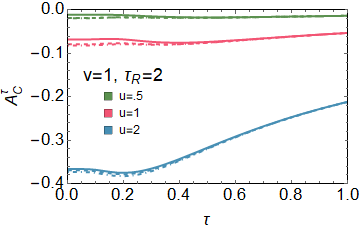}
\caption{
Correlation free energy from effective thermal potential 
for the same system as previous figures.  Solid is exact, dashed is from 
Eq. (\ref{XCcorrections}), and dot-dashed includes the further approximation of
Eq. (\ref{lambdaapprox}).
}
\label{xcfreeenergy}
\end{figure}
In 
Fig. \ref{xcfreeenergy}, we plot correlation energies 
exactly, approximately but doing the coupling-integral in
Eq. (\ref{XCcorrections}) explicitly,
and approximately but with Eq. (\ref{lambdaapprox})
to approximate the coupling-integrations.
Using Eq. (\ref{XCcorrections}) introduces small errors
for low temperatures but these quickly
diminish as temperature increases.
Interestingly, they seem no worse when correlations are stronger.
Linearizing the potential slightly worsens the results, but 
makes a smaller error than already present in Eq. (\ref{XCcorrections}). 
This error also diminishes rapidly with increasing temperature.
As noted in
Ref. \cite{SPB16}, correlation becomes a relatively smaller part of the
total free energy as temperature increases.
The majority of the contribution to the correlation free energy
is in the reference term with similar behavior in the 
correction term as seen for the total free energy, and we are making only
a small error in approximating the correction.

In this work we have presented a formally exact method for determining electronic
properties at temperature $\tau$ using a calculation at temperature $\taur$. To do so, we 
defined an effective thermal potential which yields the exact density at $\tau$
of a given system.
We have illustrated the effective thermal potential using the asymmetric Hubbard dimer. 
We have also derived an approximate formula from potential functional theory for the exchange-correlation free energy
that uses only the effective thermal potential.
We applied simple approximations to this equation to put it in a more elegant form and to make it
only require the effective thermal potential.

We conclude with suggestions for approximations and future work.
For extended matter in WDM simulations, an obvious reference potential is the uniform electron
gas with the average electronic density of the entire system.  The free energy of this
system is (relatively) well 
known\cite{BCDC13,KSDT14,SGVB15,MBBL16,GDB17}.  Then the coupling-constant integral connects local differences in
the potential from its average value.  Eq. (\ref{vtil}) would also be first tested with e.g., a zero-temperature
GGA approximation for the MKS approximation.  This yields an approximate density at $\tau$
and the corresponding HXC approximation at $\taur$.  Then the same MKS code could be used to find the
corresponding MKS potential at $\taur$, by using any one of several feedback schemes to adjust $v\s^\taur(\br)$ until
the one yielding $\tilde\n\t(\br)$ is found.  These then yield the approximation to the effective thermal 
potential to be used in an accurate quantum solver at $\taur$.  Note that one could also imagine this as the
first step in an iterative procedure in which the output approximate density at $\tau$ is used in place of the
MKS approximate density.  This would unbalance the use of DFT in the formula which might in fact worsen the
results.  Only practical calculations can tell.
Additional future tests and demonstrations of this theory include long Hubbard chains,
more complicated lattices, and atoms. These tests can further show the theory's 
applicability and accuracy.


\acknowledgments

The authors acknowledge support from the US Department of Energy (DOE), Office of Science, 
Basic Energy Sciences under Award No. DE-FG02-08ER46496. J.C.S. acknowledges
support through the NSF Graduate Research fellowship program under Award No. DGE-1321846.

\bibliography{Master,hubbard,thermal}
\label{page:end}
\end{document}